# The Aleph & Other Metaphors for Image Generation

**Gonzalo A. Ramos**
Microsoft Research
goramos@microsoft.com

**Rick Barraza**
Microsoft
ricbarr@microsoft.com

**Victor Dibia**
Microsoft Research
victordibia@microsoft.com

**Sharon Lo**
Microsoft
shlo@microsoft.com

## Abstract

In this position paper, we reflect on fictional stories dealing with the infinite and how they connect with the current, fast-evolving field of image generation models. We draw attention to how some of these literary constructs can serve as powerful metaphors for guiding human-centered design and technical thinking in the space of these emerging technologies and the experiences we build around them. We hope our provocations seed conversations about current and yet-to-be developed interactions with these emerging models in ways that may amplify human agency.

## 1 Borges, infinites and latent spaces.

The works of the writer Jorge Luis Borges often deal with fantastical and mathematical themes, and of these, notions about the infinite stand out with subtle connections to the fast-paced emergence of image generation models (IGMs) such as Dall-E 2, Stable Diffusion, and Imagen [10, 11, 12].

In *El Aleph* (1945) [2], we read about the existence of a space anomaly: "the Aleph", a point in space containing all other points. When looking at it, the protagonist of the story can see each thing in it as "infinite things, since I distinctly saw it from every angle of the universe". There are undeniable parallels between this concept and IGMs' latent spaces that have richly embedded semantic meaning for the visual spaces they encode. These spaces allow for computational variations of an image and also computational interpolations between the images themselves.

In *La Biblioteca de Babel* (1941) [3], Borges makes us reflect about infinite spaces and the limits of human knowledge. "Babel's Library" is a universe, an infinite-size structure made of interconnected hexagonal galleries, each containing exactly 640 books, each written using a script consisting of 25 different characters. Borges provokes the reader by arguing that, because of the finite nature of human language, such a library must contain a finite number of sensical books[1]. Somewhere within this vast collection of possible permutations, any story imaginable would be found, but the searching librarian is hopelessly lost among the magnitude of possibilities. There is now the provocation that IGMs may provide a computational mapping of the near-infinite possibilities of meaningful embeddings.

*El Libro de Arena* (1941) [1] can be seen as a cautionary tale about the addictive perils of having access to an infinite generator of stories. The reader of this hypothetical book is ultimately horrified by it and decides to get rid of it, not by destroying it (burning an infinite book will suffocate the world), but by hiding it in a library's basement expressing "the best place to hide a leaf is in a forest".

---

[1] This argument can be followed by thinking about constraining the size of each book to a certain number of characters and pages.

Others are rightfully better suited to speak about the legacy of Borges' writings. Instead, we draw attention to how some of his literary constructs can serve as powerful metaphors for guiding human-centered design and technical thinking in the space of IGMs.

## 2 Metaphors of the infinite as provocations.

Metaphors are powerful instruments that we use to try to make sense of what we do not know: we understand and measure the world using metaphors. Metaphors are at the center of the creative process in the fields of interaction and experience design. The desktop, the typewriter are famous examples of metaphors that still guide the ways we think about interacting with computers. Metaphors are useful until they break, and then they open the door to a new space of innovation from which new metaphors emerge.We present two provocations that we hope inspire conversations that lead to more human-centered outcomes in the context of these emerging IGMs.

### 2.1 Infinite library metaphors help us understand IGMs from a human-centered ML lens.

Libraries, generalize bookshelfs and remain a relatable, knowable place where one goes to obtain a particular book, or search for one that will satisfy a goal. This includes exploration, since surprise is a valid goal. Making the library infinite does not negate its utility, but allows for a familiar lens to the potentially infinite number of image representations (books) contained within the latent spaces of IGMs [2]. The usefulness of the metaphor stems from the familiar (human-centered) interfaces libraries have for people to interact with them: there is a librarian that can help us find what we are looking for, there is an indexing system, there are thematic hallways, etc. In contrast, some metaphors provide less defined agencies. Thinking about image generators as entities that "dream" or "hallucinate" capture their stochastic nature, but what is the interface or agency we have over dreams? The next provocation elaborates on the importance of a metaphor's interfaces.

### 2.2 These metaphors' interfaces inspire the design and development of meaningful experiences with IGMs.

Current text/image to image experiences already allude to a librarian or curator. One merely must ask (through a multimodal prompt) for a particular book, and an encoder [9] (i.e. librarian) maps that ask into a sensible location in the library. This capability is an intellectually significant departure from Borges' pessimistic hopes to index the library. Still, this does not mean that there is a systematic index of the latent space understandable to humans; it is not yet clear how to navigate the library's hallways in semantically sensible ways. Prompt engineering is an indirect (chaotic even) way to navigate the latent space. The arcane nature of prompt articulation has already launched marketplaces [8] where prompt experts map, trade, and gatekeep access to safe destinations in the library. A library metaphor inspires us to think about experiences where people (not prompt whisperers) can explore the library's hallways with a semblance of agency that can include, but go beyond prompt trial and error, or predefined dimensions of style such as artist, camera [16], or generation parameters such as notions of temperature, guidance, seeds, and diffusion steps. There are promising ideas ahead. Text inversion [5] hints at potential solutions to define concepts that allow one to explore the latent space in personal, meaningful ways. Current advances in simulation environments used to safely train autonomous agents [13] or face detectors [15] hint at the possibility of defining training sets that include variations along semantically useful concepts. Creative communities are already combining components in pipelines that allow them to craft sequences of semantically and temporally coherent images [7, 4, 6]. An Aleph has less defined interfaces, but surprise is sometimes a goal in itself. An example of this is *Quantum Mirror* [14], an Aleph that allows the viewer to take a glance at the many variations of a point in space. It is essential to support efforts in UX design, tooling, and learning algorithms that enable people navigating with intention and agency from one image to the next.

## 3 Closing thoughts: other metaphors and perspectives.

Other metaphors emerge as we witness the emergence of IGMs and how they are being used. IGMs can be seen as a *Material*, a *Tool*, and in particular, as a communication and artistic *Medium*. Each of these metaphors brings their own set of affordances that guide how to understand and engage with them. We encourage readers to take these provocations as a starting point to think about how we present technology to others, how we expect them to be used (for good or ill), and what affordances we include by design in our technologies.

---

[2]There can be many infinite libraries, depending on how one trains the generation model.